\documentclass[a4paper]{article}
\usepackage{graphicx}

\begin{document}

\begin{center}
{\bf \Large
Magnetism of frustrated regular networks}
\bigskip

{\large
Anna Ma\'nka-Kraso\'n$^{\dag}$ and Krzysztof Ku{\l}akowski$^{\ddag}$
}
\bigskip

{\em
Faculty of Physics and Applied Computer Science,
AGH University of Science and Technology,
al. Mickiewicza 30, PL-30059 Krak\'ow, Poland\\

}

\bigskip
$^{\dag}${\tt manka@novell.ftj.agh.edu.pl}
$^\ddag${\tt kulakowski@novell.ftj.agh.edu.pl}

\bigskip
\today
\end{center}

\begin{abstract}
We consider a regular random network where each node has exactly three neighbours. 
Ising spins at the network nodes interact antiferromagnetically along the links.
The clustering coefficient $C$ is tuned from zero to 1/3 by adding new links. At 
the same time, the density of geometrically frustrated links increases. We calculate 
the magnetic specific heat, the spin susceptibility and the Edwards-Anderson order 
parameter $q$ by means of the heat-bath Monte Carlo simulations. The aim is the 
transition temperature $T_x$ dependence on the clustering coefficient $C$.
The results are compared with the predictions of the Bethe approximation. At $C=0$, 
the network is bipartite and the low temperature phase is antiferromagnetic. 
When $C$ increases, the critical temperature falls down towards the values which 
are close to the theoretical predictions for the spin-glass phase. 

\end{abstract}

\noindent
{\em PACS numbers: 75.30.Kz, 64.60.aq, 05.10.Ln} 

\noindent
{\em Keywords: antiferromagnetism, spin glass, disorder, frustration, regular random networks}  

\bigskip

\section{Introduction}

Statistical mechanics of random networks gained recently many applications in interdisciplinary sciences. The list of references is already very rich; 
in almost each year new monographies appear \cite{lkd,dog,hand,pasat,dur,card}. In theory, the promising challenge is to investigate collective 
phenomena in networks \cite{dog1}. More than often it is useful to decorate nodes $i$ with additional variables, as Ising spins $S_i=\pm 1$.
Properties of such networks are of interest for theory of computations, as inference problems, but also for the theory of disordered magnetic systems. 
Here we are interested in the spin-glass phase of the small-world random networks. \\

For trees, where closed loops are absent, there is a consistent theory of Ising magnetism, i.e. the Bethe theory \cite{bethejap,bax,mooij}. Once the loops
appear in the system, the Bethe theory becomes an approximation. In our system, small loops are introduced when we enhance the clustering coefficient \cite{hol,my}. Still, the results of the Bethe theory are useful as a point of reference. As it was demonstrated in \cite{my}, the accordance of the numerical results with this theory was deteriorated when $C$ increases. On the other hand, there is at least one periodic two-dimensional Ising lattice
when all bonds are antiferromagnetic, the density of frustration is high and the transition temperature is positive; this is one of the Archimedean 
lattices \cite{kraw}. The ground state of this system is highly frustrated, but at least some of the energy barriers between different states remain finite. Although at zero field the net magnetization is zero, there is no disorder in the system; therefore it is hard to speak about the spin-glass phase. Then, the $(3,12^2$) Archimedean lattice can be compared to our system in the case where $C$ is maximal.\\

Recently we investigated the transition from the paramagnetic to the spin-glass phase in the random Erd\"os-R\'enyi network with enhanced clustering coefficient $C$ \cite{my}. The enhancement is introduced by adding new links between neighbours of the same sites \cite{hol}. In this way, $C$ varied from almost zero to about 0.3. The main result was the transition temperature $T_{SG}$ dependence on the  clustering coefficient $C$. However, in the numerical plots the transition was partially hidden by the contribution of spins which could flip with zero energy cost. That is why here we investigate the same transition in the regular random network, where the degree of each node is an odd number - here it is equal to three. In such systems no spin can flip at zero energy cost in zero external magnetic field. \\

In the next section we show the method and the numerical results. In Section III the transition temperature $T_x$ is compared with the predictions of the Bethe theory. Short discussion closes the text.\\

\begin{figure}
\begin{center}
\includegraphics[scale=0.45,angle=-90]{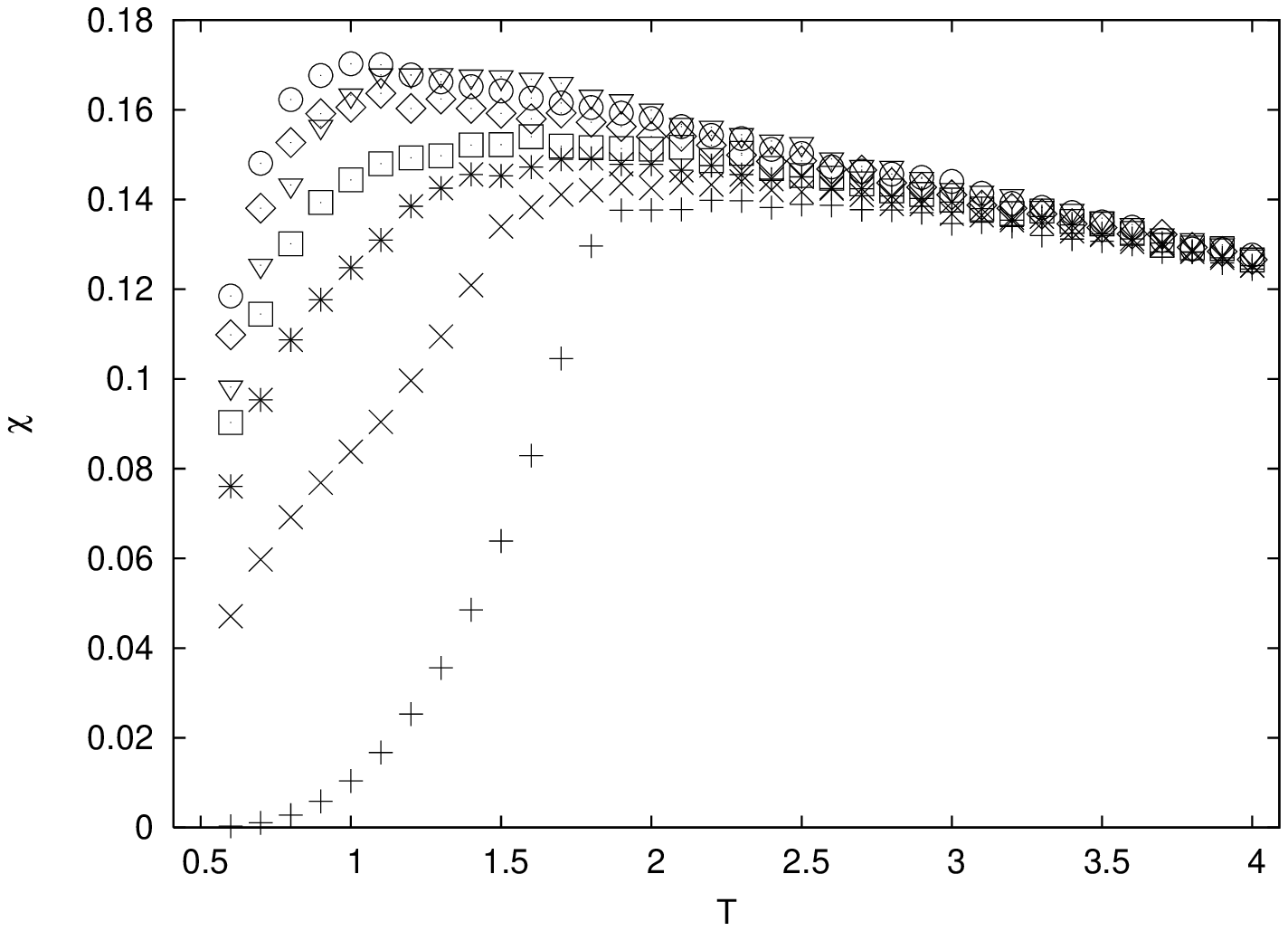}
\end{center}
\caption{The magnetic susceptibility $\chi$ calculated from the variance of the magnetization against temperature $T$, for different values of the 
clustering coefficient $C$=0.0, 0.08, 0.14, 0.18, 0.25, 0.29 and 0.33. With increasing $C$, the maximum of $\chi$ shifts to low temperatures till $C$=0.18, then remains approximately constant} 
\end{figure}

\section{Method and results}

Initial form of the constructed network is prepared as an even number $N_0$ of unlinked nodes. The system is divided into two equal parts. Three 'dangling' links are assigned to each node. Then, the links of one part of nodes are randomly joint to links of the second part. In this way, initially the system is bipartite, with the clustering coefficient $C$ equal to zero provided that the number of nodes is large enough. Next step is performed for each node with the probability $p$: the node is substituted by a group of three nodes, mutually linked. After this substitution, three previous neighbours of the node are linked to the nodes of the group. In this way, the degree of each node remains exactly three. As the Ising interaction is exclusively antiferromagnetic, the frustration is purely geometrical. When $C=0$, there is no frustration. The density of frustrated bonds increases with $C$. 
If $p=1$, the clustering is maximal and for each node, two out of its three neigbours are linked to each other; then $C=1/3$. The final number of nodes
is $N=N_0(1+2p)$; for each $p$ the amount $N_0$ is chosen as to get approximately the same $N$.\\

The heat-bath Monte Carlo algorithm is applied to investigate the magnetic properties: the susceptibility $\chi$, the specific heat $C_v$ and the Edwards-Anderson order parameter $q$. $\chi$ can be calculated from the field derivative of the magnetization $m(h)$ or from the variance of the spectrum of $m(0)$. Similarly $C_v$ can be obtained from the temperature derivative of energy or from the variance of the energy fluctuations. The parameter $q$
is calculated from the formula \cite{by}

\begin{equation}
q=\frac{1}{N}\sum_{i=1}^N\Big[\frac{1}{\tau}\sum_{t=0}^\tau s_i(t)\Big]^2
\end{equation}
The antiferromagnetic exchange integral $J$ is set to -1. As a rule, the calculations are performed for $N$ equal at least $9\times 10^5$ spins.  The time of calculation - after a necessary transient - was usually $10^5$ steps, where one step is equivalent to probing $N$ spins in random order. The results are limited to temperature $T>0.5$, where the transient time for relaxation of energy was smaller than $10^3$ timesteps.\\

\begin{figure}
\begin{center}
\includegraphics[scale=0.45,angle=-90]{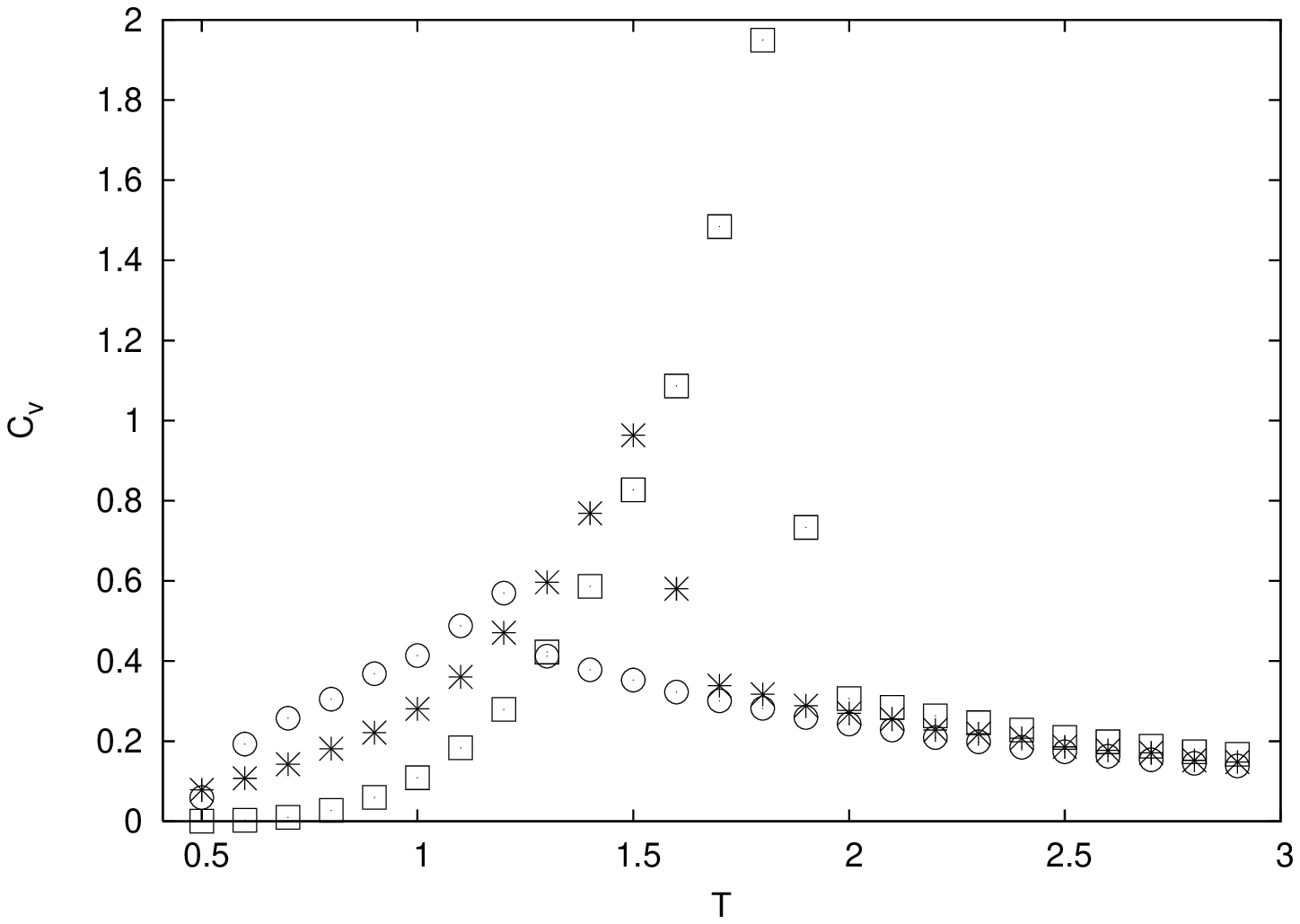}
\end{center}
\caption{The magnetic specific heat $C_v$ against temperature $T$ for different values of the clustering coefficient $C$=0.0 (squares), 0.08 (stars) and 0.14 (circles). With increasing $C$, the maximum of $C_v$ shifts monotonously to low temperatures.} 
\end{figure}

In Fig. 1 we show the magnetic susceptibilities $\chi (T)$ for different values of the clustering coefficient $C$. The temperature is expressed in energy units $-J$. The finite size effect is checked for $C=0$ and $C=0.33$; the differences between $N=9\times10^3$ and $N=9\times10^5$ are invisible. As a rule, the plots obtained from the field derivative of the magnetization coincide with those from the variance of the magnetization at zero field above the transition temperature. Below $T_x$, the curves split for intermediate values of $C$ (between 0.1 and 0.29). The plots for the 
magnetic specific heat obtained from the variance of energy (Figs. 2 and 3) and from its thermal derivative coincide in most cases, even below $T_x$ and for intermediate values of $C$. However, the data from the maxima of the specific heat are known to provide an evaluation of the upper limit of the transition temperature $T_x$ rather than $T_x$ itself. The results on $T_x$ obtained by different methods are compared in Fig. 5.  As we see, there is a systematic split between $T_x$ obtained from $\chi $ and $T_x$ obtained from $C_v$, but the character of the curve $T_x(C)$ is preserved. In Fig. 4 we show the data on $T_x$ obtained from the thermal dependence of the Edwards-Anderson order parameter $q$. These data also show a decrease of $T_x$ with $C$. However, this decrease persists till $T=0.5$, where our numerical results are less reliable.\\

\section{Discussion}

For $C=0$ the network is bipartite. Although the interaction is antiferromagnetic, the system is equivalent to a ferromagnet. To see this, we have to 
flip spins at half of nodes and to invert simultaneously the sign of all exchange integrals $J$ from -1 to +1. The formula for the Curie temperature ($J=+1$) for the regular Bethe lattice with the coordination number $k$ is \cite{bax}

\begin{equation}
T_c=\frac{2J}{ln\frac{k}{k-2}}
\end{equation}
In this case this formula can be applied equally well to the N\'eel temperature ($J=-1$). In both cases the number of loops in the system is exactly zero. 

\begin{figure}
\begin{center}
\includegraphics[scale=0.45,angle=-90]{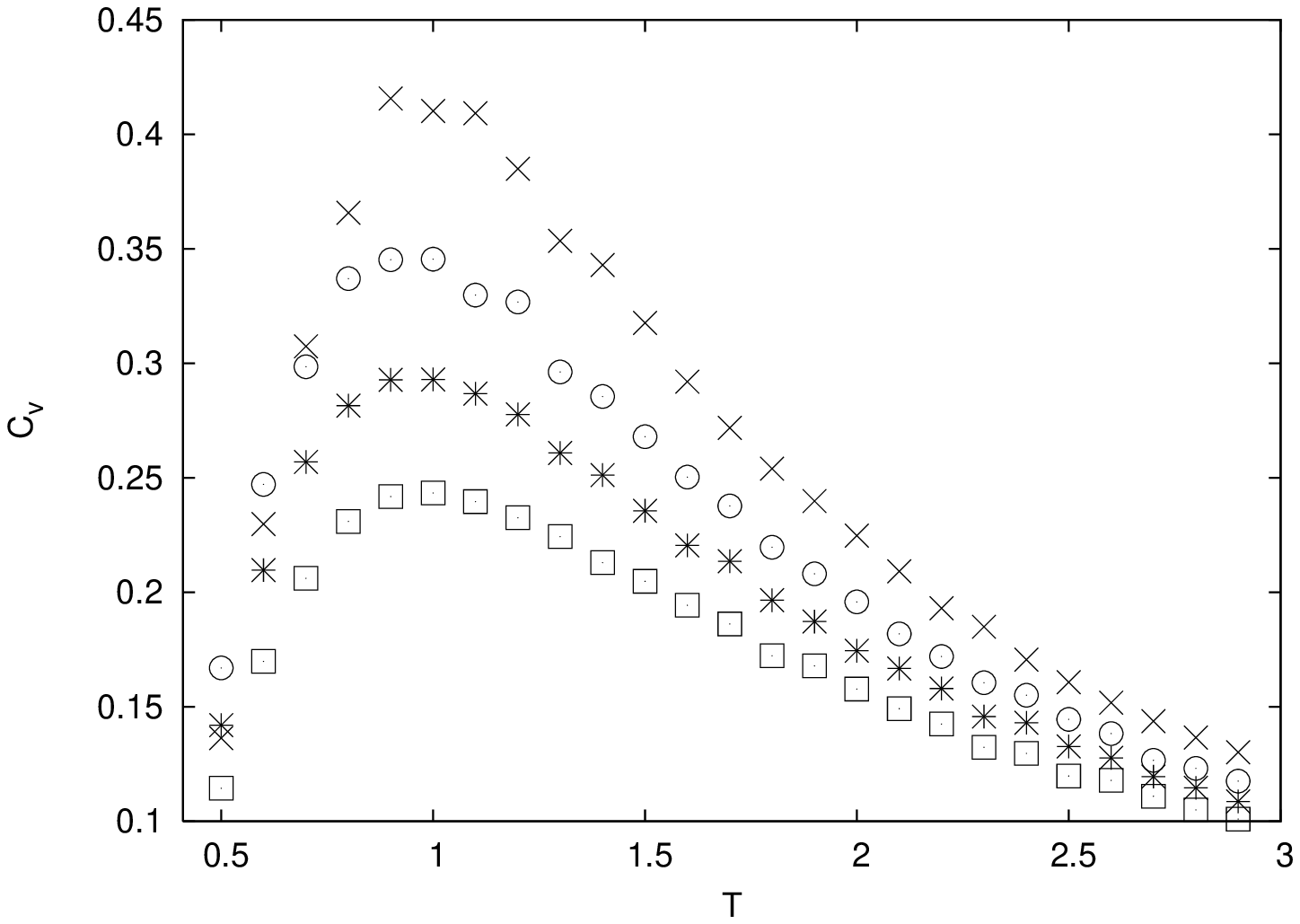}
\end{center}
\caption{The magnetic specific heat $C_v$ against temperature $T$ for different values of the clustering coefficient $C$= 0.18, 0.25, 0.29 and 0.33. With increasing $C$, the height of the maximum of $C_v$ decreases.} 
\end{figure}

For $C>0$ the system contains loops and it is no more equivalent to a ferromagnet. On the other hand, the network is no more a regular Bethe lattice. The 
number of second neighbours decreases gradually with $p$ from 6 when $p=0$ to 4 when $p=1$. The latter case is, however, in some sense more regular than 
the one with smaller $p$, because all nodes have again the same number of the second neighbours. Then the N\'eel temperature ($J=-1$) between the paramagnetic and the antiferromagnetic phase is to be found from

\begin{equation}
T_N=\frac{-2J}{ln\frac{B+1}{B-1}}
\end{equation}
and the transition temperature between the paramagnetic and the spin-glass phase -- from

\begin{equation}
T_{SG}=\frac{-2J}{ln\frac{\sqrt{B}+1}{\sqrt{B}-1}}
\end{equation}
where $B$ is the average branching parameter, i.e. $B=z_2/z_1$, where $z_1$ ($z_2$) is the average number of first (second) neighbours \cite{dog1}.\\

Although the data from the specific heat fit better to the Bethe theory, it is known that the position of the maximum of $C_v$ is rather the upper bound of the transition temperature than this temperature itself.\\

\begin{figure}
\begin{center}
\includegraphics[scale=0.45,angle=-90]{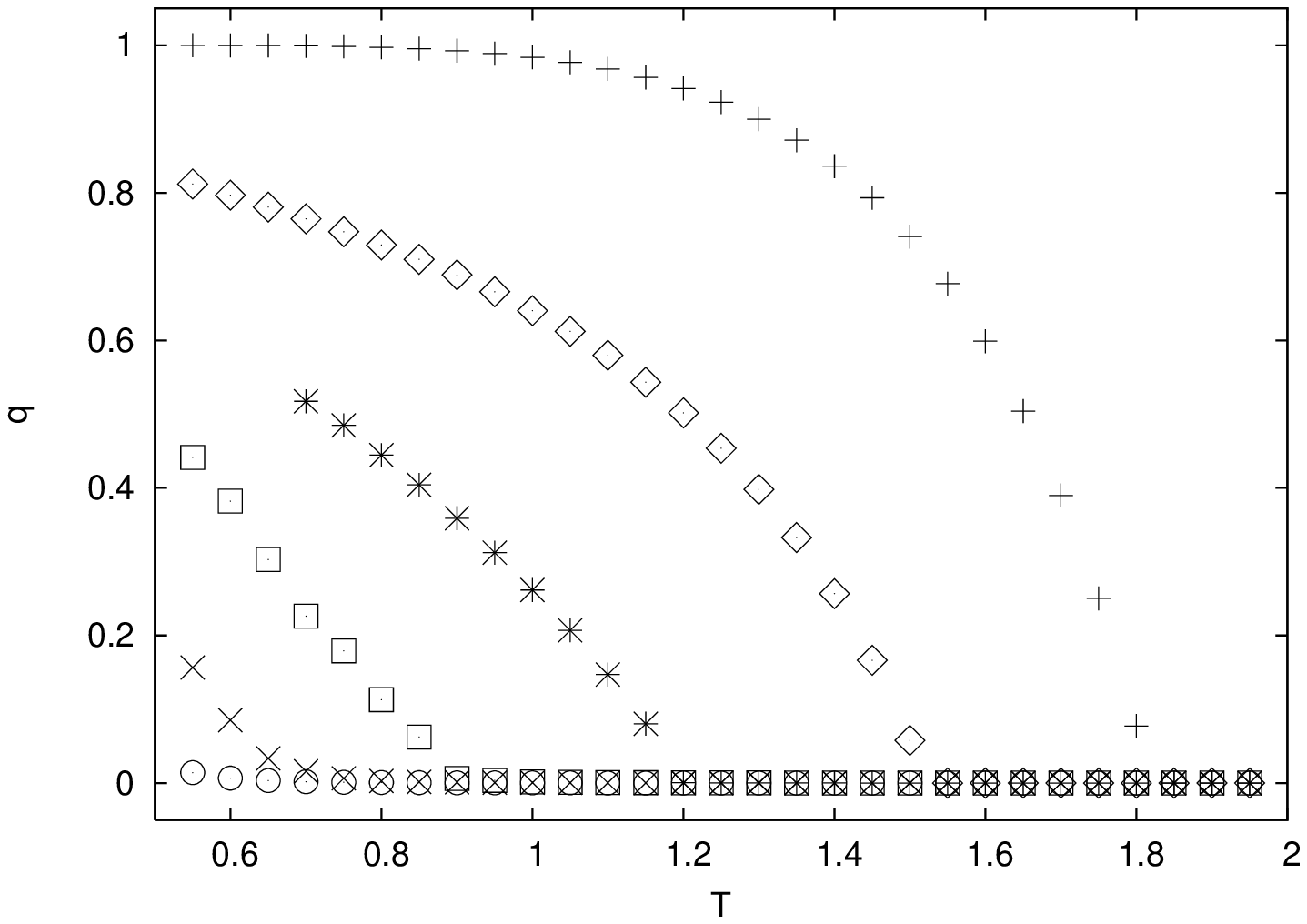}
\end{center}
\caption{The Edwards-Anderson order parameter $q$ against temperature $T$ for different values of the clustering coefficient $C$=0.0, 0.08, 0.14, 0.18, 0.25. With increasing $C$, the temperature $T_x$ where $q$ vanishes shifts monotonously to low temperatures.} 
\end{figure}

The numerical results indicate, that for $C=0$ the low-temperature phase is antiferromagnetic. They also suggest that for $C>0.1$ and $T<T_x$ the system becomes a spin glass. However, the identity of this phase remains not clear. For $C=1/3$ the system can be compared to the  $(3,12^2)$ Archimedean lattice \cite{kraw} with random rewiring. We can expect that the ground state is degenerated in the same way as in this Archimedean lattice. However, the condition is that the rewiring does not destroy the small triangles introduced when $C$ is enhanced. In this way the frustration is not altered by rewiring: the frustration remains local. Then, there is frustration and disorder, but the disorder does not influence the frustration. On the other hand, the energy barriers between the local energy minima remain finite, because the system can move from one minimum to another by flipping a finite amount of spins. Although the system behaves as a spin glass, some ingredients of this mysterious phase are missing.\\

\begin{figure}
\begin{center}
\includegraphics[scale=0.45,angle=-90]{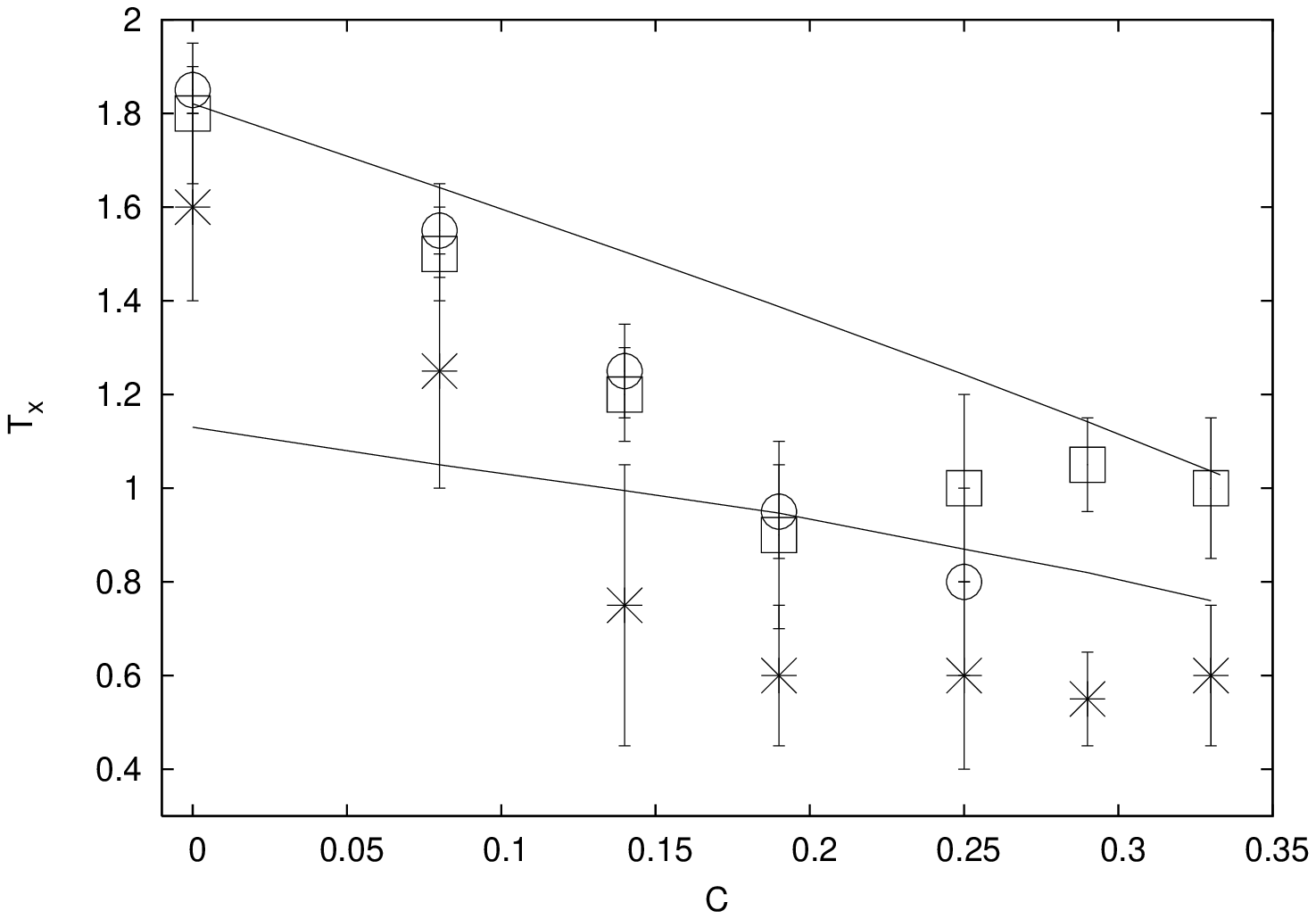}
\end{center}
\caption{The critical temperature $T_x$ against the clustering coefficient $C$, calculated numerically from the inflection point of the magnetic susceptibility $\chi$ (stars), the maximum of the specific heat $C_v$ (squares) and the Edwards-Anderson order parameter $q$ (circles). Two lines
mark theoretical values of $T_x$ for the para-antiferromagnetic phase transition (Eq. 3, upper line) and the paramagnetic- spin glass transition
(Eq. 4, lower line).} 
\end{figure}

To summarize, our intention was to compare the numerical results on the transition temperature $T_x$ with the results of the Bethe approximation. In Ref. \cite{my}, distinct departures have been found between the Bethe theory and the numerical experiment. Here the departure is smaller and quantitative rather than qualitative. As a rule, the transition temperature is overestimated by the Bethe approximation. Most important difference between the system
considered in this text and the system discussed in Ref. \cite{my} is the degree distribution. This suggests, that the qualitative departure of the results of Ref. \cite{my} from those of the Bethe approximation is due to the variance of the degree distribution. \\

\bigskip

{\bf Acknowledgements.} The calculations were performed in the ACK Cyfronet, Cracow, grants No. MNiSW /SGI3700 /AGH /030/ 2007 and MNiSW /SGI3700 /AGH /031/ 2007. 

\end{document}